\pdfoutput=1
\documentclass[12pt]{iopart}
\usepackage{graphicx}
\usepackage{bm}

\begin{document}

\article{}{Scale free networks by preferential depletion}

\author{Christian M. Schneider$^1$, Lucilla de Arcangelis$^{1,2}$ and Hans J. Herrmann$^{1,3}$}
\address{$^1$ Computational Physics, IfB, ETH Zurich, Schafmattstrasse 6, 8093 Zurich, Switzerland}
\address{$^2$ Dept. of Information Engineering and CNISM, Second University of Naples, 81031 Aversa (CE), Italy}
\address{$^3$ Departamento de F\'{\i}sica, Universidade Federal do Cear\'a, 60451-970 Fortaleza, Cear\'a, Brazil}

\ead{schnechr@ethz.ch, lucilla@na.infn.it, hans@ifb.baug.ethz.ch}

\begin{abstract}
We show that not only preferential attachment but also preferential depletion leads to scale-free networks. The resulting degree distribution exponents is typically less than two (5/3) as opposed to the case of the growth models studied before where the exponents are larger. Our approach applies in particular to biological networks where in fact we find interesting agreement with experimental measurements. We investigate the most important properties characterizing these networks, as the cluster size distribution, the average shortest path and the clustering coefficient.
\end{abstract}

 \pacs{64.60.aq, 
       89.75.Da, 
       89.75.Fb, 
       89.75.Hc 
       }


\maketitle
\section{Introduction}
From technological networks, like the Internet \cite{faloutsos99}, to social contact networks, like sexual contacts \cite{liljeros01} or friendship networks \cite{barabasi01}, to biological ones \cite{chung03}, networks are emerging in a
variety of different fields. Often these networks are scale free, which means that the distribution of connections per node follows a power law. The understanding of the origin of this power-law can give insights in the network evolution and role.\\
The first explanation of the power-law degree distribution was proposed by Barabasi and Albert \cite{barabasi99}. They identified two mechanisms, growth and preferential attachment, as the main ingredients leading to scale-free networks. An increasing number of nodes and a constant number of new connections between new and old nodes, created preferentially with high degree nodes, lead to a power-law distribution.\\
After this seminal work, other models with additional features as accelerating growth \cite{dorogovtsev01}, aging \cite{dorogovtsev00b}, initial attractiveness \cite{dorogovtsev00a} or fitness \cite{bianconi01} have been developed. In the case of accelerating growth, additional edges are added between high degree nodes, whereas in the model with aging, old nodes are less likely to create new connections. Moreover, the influence of the connectivity can be reduced by a constant initial attractiveness or even replaced by a fitness function \cite{bianconi01}. All these models have the same basic ingredients: growth and preferential attachment. Even replacing preferential attachment by similar mechanisms, the evolution towards scale-free networks remains possible. For example, nodes and edges can be copied \cite{kumar00}, edges redirected \cite{krapivsky01}, walkers can create additional connections \cite{vazquez00} or deterministic rules implemented \cite{andrade05}. In all these cases the power-law behavior is based on the creation of new connections or nodes. In Table \ref{tab1} the exponents $\gamma$ of the degree distribution for a variety of scale-free models are listed.\\ 
\begin{table}
 \centering\small
 \begin{tabular}{l|c|c}
 Model or Network & $\gamma$ & Ref.\\\hline
 Linear growth & $3$ & \cite{barabasi99}\\
 Accelerating growth & $1.5$ and $3$* & \cite{dorogovtsev01}\\
 Aging & $2 - \infty$ & \cite{dorogovtsev00b}\\
 Initial attractiveness & $2 - \infty$ & \cite{dorogovtsev00a}\\
 Fitness & $k^{-1-C}/\ln k$**& \cite{bianconi01}\\
 Copying with probab. p & $(2 - p)/(1 - p)$ & \cite{kumar00}\\
 Redirection with probab. p & $1 + 1/p$ & \cite{krapivsky01}\\
 Walker with probab. p & $2$ for $p > 0.4$& \cite{vazquez00}\\\hline
 Internet & $2.5$ & \cite{faloutsos99}\\
 Movie actors & $2.3$ & \cite{barabasi99}\\
 Co-authors & $2.5$ & \cite{barabasi01}\\
 Sexual contacts & $3.4$ & \cite{liljeros01}\\
 Citation & $3$ & \cite{redner98}\\\hline
 E.coli metabolic network & 1.7 & \cite{Friedman}\\
 Gene expression data & 1.5 & \cite{DeRisi,Spellman}\\
 Gene functional interactions & 1.6 & \cite{Gu}\\
 Combined-AP/MS - S.cerevisiae & 1.5 & \cite{Yeast}\\
 Integrated Network - C.elegans & 1.2 & \cite{Yeast}\\
 Interolog - C.elegans & 1.5 & \cite{Yeast}\\
 Genetic interaction network & 1.7 & \cite{Tang}
\end{tabular}
 \caption{Exponents of the degree distribution for different scale-free models and
some social and biological networks. *two regimes; **$C$ depends on fitness
distribution}
 \label{tab1}
\end{table}
In all of these growing network models, the exponent $\gamma$ of the degree
distribution is larger than two, with the smallest value typically obtained for
extreme values of the parameters of the respective model. They are appropriate to
describe social networks, since typical social scale-free networks have a
$\gamma$-value between two and three, however biological networks are characterized
by significantly smaller values of $\gamma$ \cite{chung03}.

\begin{figure}
\includegraphics[width=16cm,angle = 0]{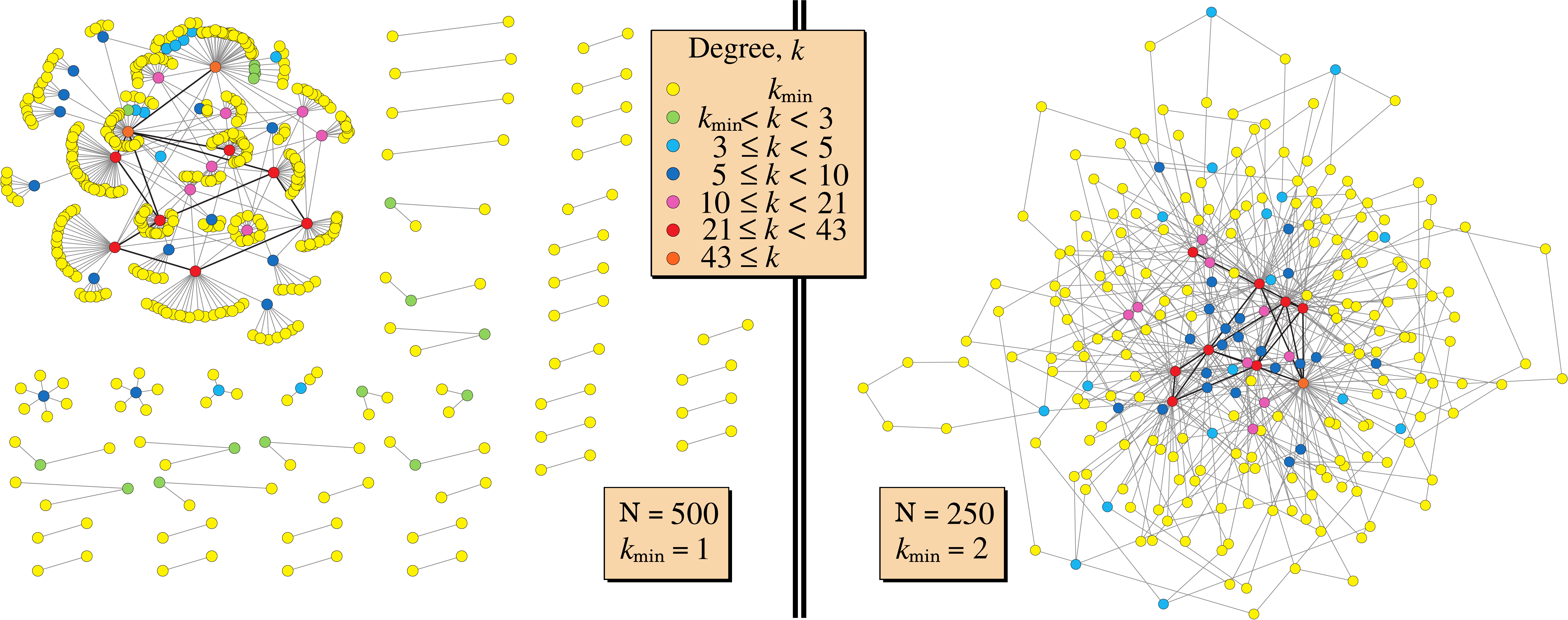}
\caption{(Colors online) Examples of scale-free networks created by depletion with $\alpha = 2$, $k_\mathrm{min} = 1$ and $N = 500$ (left side) and $\alpha = 2$, $k_\mathrm{min} = 2$ and $N = 250$ (right side). The color code represents the degree of each site on a logarithmic scale and the connections between the hubs are highlighted. For $k_\mathrm{min} = 1$ the network consists of a large number of small clusters, whereas for larger values of $k_\mathrm{min}$ only a single cluster is present. In both cases hubs survive the depletion process\cite{Pajek}.}
\label{fig:1}
\end{figure}

\section{Depletion model}
Here we show that the preferential depletion of links, keeping the number of nodes
constant, is a mechanism to generate scale-free networks with a small $\gamma$
value. The idea is inspired by the plastic adaptation of neuronal networks in the
brain. Indeed, plasticity is one of the most astonishing properties of the brain,
occurring mostly during development and learning \cite{plast,plast2}, and can be
defined as the ability to modify the structural and functional properties of
synapses. Modifications in the strength of synapses are thought to underlie memory
and learning. Progressive depression of synaptic strength can lead to "pruning",
i.e. removal of connections. This remodeling of synapses mimics the fine tuning of
wiring that occurs during "critical periods" in the developing brain, when neuronal
activity can modify the synaptic circuitry, once the basic patterns of brain wiring
are established \cite{plast2}. Experimental measurements of the functionality
network in human adults have evidenced that this is scale free \cite{egui}.
Functional magnetic resonance imaging has indeed shown that this network has
universal scale-free properties: it exhibits a scaling behavior for the out-degree
distribution with an exponent $\gamma\simeq 2.0$, independent of the different tasks
performed by the patient. It is interesting to notice that the $\gamma$ exponent
found in experiments is smaller than the typical value found for non-biological
networks (Table 1). Moreover, neuronal network models \cite{Lucilla,Lucilla2} have 
evidenced that the presence of highly connected nodes is crucial for learning 
\cite{Lucilla3}.

To implement this idea, we develop the following algorithm. The model starts with a fully connected network with $N$ nodes. The evolution then removes edges according to the following rules:
\begin{itemize}
 \item Choose randomly a node $i$.
 \item Choose one of the edges of node $i$, $e_{ij}$, according to the 
probability $p_{i,j}$ and remove it. The probability $p_{i,j}$ is determined by the degrees $k_j$ of the neighbors $j$ of the node $i$:
\begin{eqnarray}
 p_{i,j} = \frac{p_j}{N_i}
\end{eqnarray}
where $p_j$ is
\begin{equation}
X=\cases{k_j^{-\alpha}&for $k_j > k_\mathrm{min}$\\0&otherwise\\}
\end{equation}
and $N_i$ the normalization
\begin{eqnarray}
 N_i = \sum_{l=0}^{l=k_i}{p_l}.
\end{eqnarray}
\item Repeat this procedure until the number of edges $M$ is equal to the number of nodes $N$ times the minimal degree $k_\mathrm{min}$.
\end{itemize}
The depletion model introduces two free parameters, $\alpha >0$ and $k_\mathrm{min}$, which control the morphology of the network and the scaling behavior of the degree distribution. Eq.(1) and (2) imply that the smaller the number of connections of node $j$, the higher is the probability to remove the edge $e_{ij}$, i.e. {\it the poor get poorer}. The value $\alpha =0$ corresponds to the case of random depletion.

\section{Numerical results}
The different morphologies of two typical networks created by the depletion algorithm for $k_\mathrm{min} = 1$ and 2 are shown in Fig.\ref{fig:1}. A change in the minimal degree $k_\mathrm{min}$ affects the structure of the network significantly. For $k_\mathrm{min} = 1$, we observe a large number of isolated small clusters and one large cluster, where for larger $k_\mathrm{min}$ only one single cluster is produced. In the case $k_\mathrm{min} = 1$, all small clusters have the minimal allowed number of edges $M_\mathrm{min}$ = $S - 1$, where $S$ is the number of nodes in a cluster. Furthermore, the giant cluster looks like a tree with a few hubs, each one connected to other hubs.\\
\begin{figure*}
 \includegraphics[width=4.7cm,angle = -90]{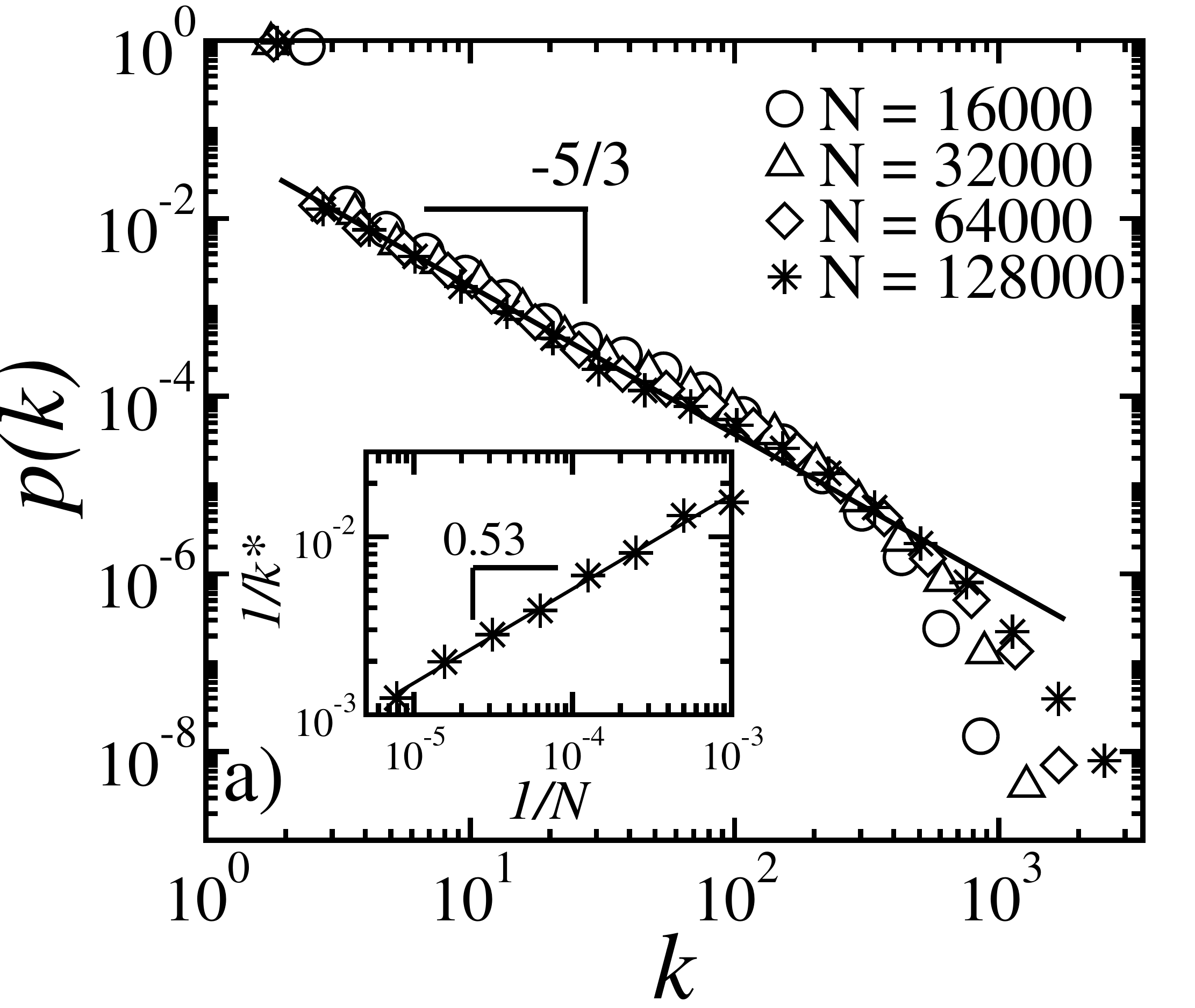}\hspace*{-0.1cm}
 \includegraphics[width=4.7cm,angle = -90]{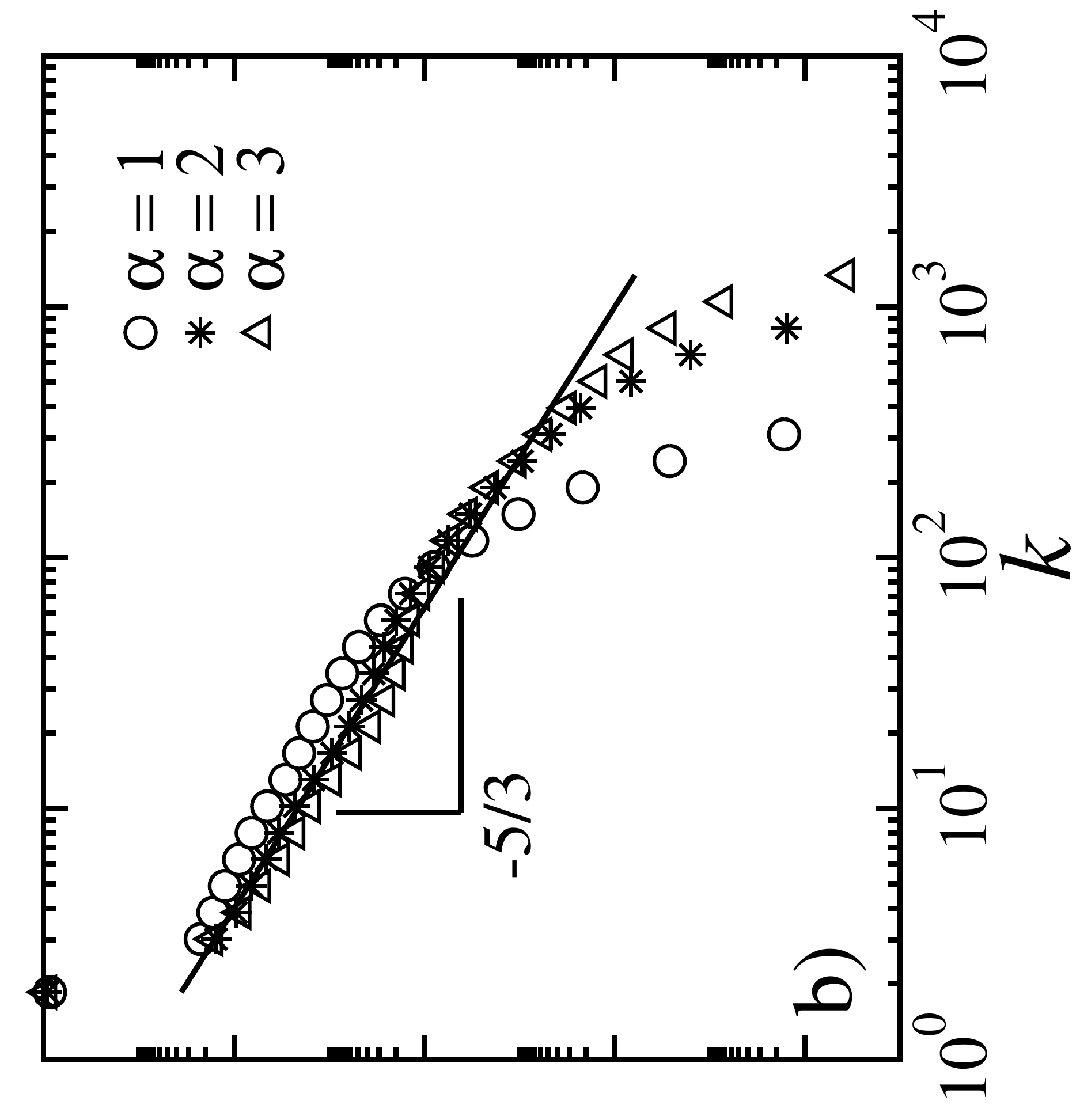}\hspace*{-0.1cm}
 \includegraphics[width=4.7cm,angle = -90]{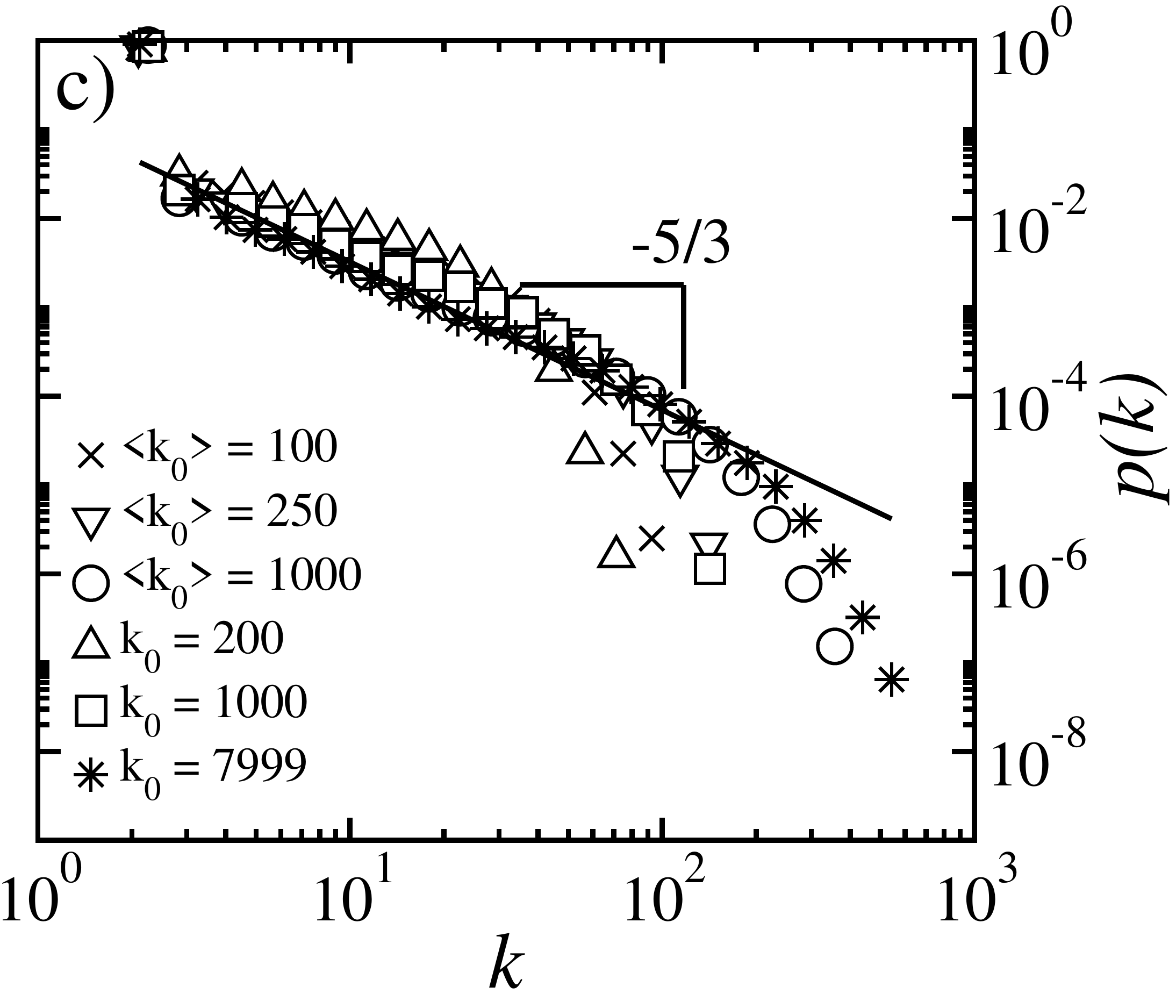}
 \caption{a) The degree distribution $p(k)$ of the depletion model with $\alpha = 2$
and $k_\mathrm{min} = 2$ for different system sizes $N$. In the inset we show the size 
dependence $1/N$ of the inverse of the crossover point $k^*$ between exponential and
power-law regime. b) The degree distribution $p(k)$ of the depletion model 
with $N = 16000$ and $k_\mathrm{min} = 2$ for different $\alpha$ values. c) The degree
distribution $p(k)$ of the depletion model with $\alpha = 2$, $k_\mathrm{min} = 2$
and $N = 8000$ for different initial conditions. The larger the average initial
degree of a regular lattice or a random network, the larger is the power-law regime
of the final network.}
\label{fig:2}
\end{figure*}
To characterize the level of connectivity of the network, we investigate the node
degree distribution (Fig. \ref{fig:2}). The value of the parameter $\alpha$ controls
not only the maximal degree, but the extension of the scaling behavior of the
distribution. Only for $\alpha \approx 2$, the algorithm creates asymptotically
scale-free networks. For $\alpha < 2$ the range of the power-law behavior is limited
by an exponential decay that sets in at intermediate $k$, the sooner the smaller the
value of $\alpha$. For $\alpha = 0$ the degree distribution is a pure exponential.
For values of $\alpha$ increasing beyond $2$ the power-law regime decreases, the
effective $\gamma$ exponent increases and the distribution develops a small bump at
intermediate $k$. (Fig. \ref{fig:2}b)\\
In contrast, the parameter $k_\mathrm{min}$ has no major impact on the overall scale-free behavior. It controls the minimal degree of the network, however the degree distributions of the entire network with the same parameters, but different $k_\mathrm{min}$ are similar. In Fig. \ref{fig:2}a the scaling of the degree distribution of our depletion model with network size $N$ is shown. The best fit is $p(k) \sim k^{-5/3}$ which represents the asymptotic behavior for infinitely large networks. This value of the exponent $\gamma$ is thus smaller than for any other scale-free model network. In the inset we see the growth of the power law regime with the size of the network. The best fit for the crossover point $k^*$ follows a power-law $k^* \sim N^{0.53}$, showing that indeed our network is asymptotically scale-free.\\
Let us next question if a fully connected initial configuration is necessary to generate scale-free networks. Therefore we test three different initial networks. First, we start from a random lattice where each node is connected with $k_0$ neighbors \cite{randomRegular}. Second, we start from a random network with a large average degree $\langle k_0 \rangle \gg 1$ \cite{Erdos}. Finally, we test a lattice with a random initial degree configuration, where the degrees are chosen from a uniform distribution, $k_0 \in \lbrack k_\mathrm{min}, \dots, N - 1 \rbrack$. For the first two initial conditions, the degree distributions for the final depleted networks show power-law regime, whose extension increases with increasing initial average degree $\langle k_0 \rangle$. Conversely, the third initial condition is not sufficient to generate a degree distribution with power-law behavior, although the starting network has a high average degree $\langle k_0 \rangle \approx \frac{N - 1}{2}$. These results, shown in Fig. \ref{fig:2}c, indicate that a dense and uniform initial network is necessary to create scale-free networks with our depletion algorithm.\\
\begin{figure}\centering
 \includegraphics[width=6.5cm,angle = -90]{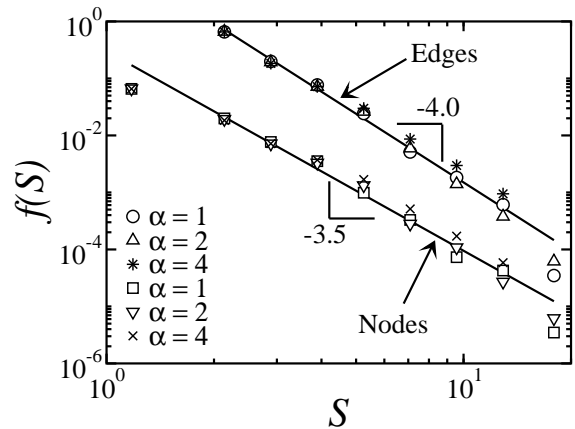}
 \caption{Frequency $f(S)$ to find a cluster with a given number of nodes $S$(bottom) or edges
$S$(top) in networks with $N = 4000$ and $k_\mathrm{min} = 1$ for different $\alpha$
values. Bottom data sets for nodes are shifted vertically by a factor of $10$ for
better visibility.}
\label{fig:5}
\end{figure}
Starting again with a fully connected network we now analyse the cluster size distribution in the case where the algorithm generates a disconnected structure. In Fig. \ref{fig:5} the frequency $f$ to find a cluster with a given number of nodes $S_N$ or edges $S_M$ is shown for the depletion model with $k_\mathrm{min} = 1$. The probability to find a cluster of given size follows approximately power-laws $f(S) \sim S^{-\beta}$ with $\beta$ around $3.5$ for nodes and $4$ for edges only weakly dependent on the parameter $\alpha$.\\
\begin{figure}\centering
 \includegraphics[width=6.5cm,angle = -90]{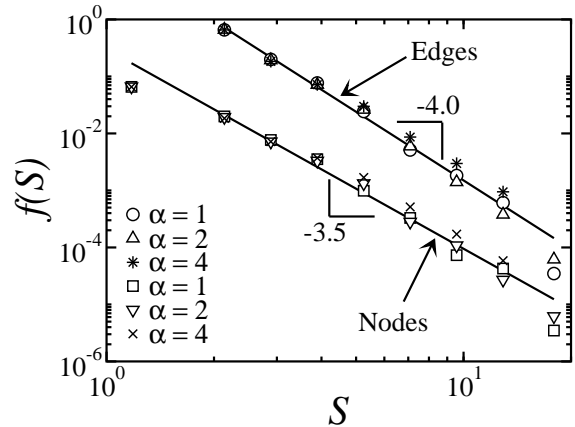}
 \caption{The size dependence of the average shortest path length $\langle l
\rangle$ for the depletion model with $\alpha = 2$ and different $k_\mathrm{min}$
values. $\langle l \rangle$ follows the double logarithmic behavior $l \propto \ln
(\ln (0.02 N))$ for $k_\mathrm{min}= 2$ and $l \propto \ln (\ln (0.21 N))$ for
$k_\mathrm{min}= 4$ respectively. Inset: The size dependence of the average
clustering coefficient $\overline{C}$ for the depletion model for both values of
$k_\mathrm{min}$. The best fit for the clustering coefficient follows a power-law
$\overline{C} \propto N^{-0.7}$ and $\overline{C} \propto N^{-0.24}$ for 
$k_\mathrm{min} = 1$ and $k_\mathrm{min} \ge 2$, respectively.}
\label{fig:6}
\end{figure}
In recent years small world properties \cite{watts98} of natural networks have emerged in different fields. These properties do not necessarily concern the degree distribution, the nodes in fact may all have a degree close to an average value. They rather characterize the level of clustering and average distance between nodes. In order to determine if the networks generated by the depletion model are small world, we evaluate the average clustering coefficient and the shortest path length (Fig. \ref{fig:6}) for $k_\mathrm{min} = 2$ and $k_\mathrm{min} = 4$. The clustering coefficient $\bar C$ is defined as the ratio of the number of observed triangles over the number of possible triangles in the network. The best fit for the average clustering coefficient $\bar C$ for networks with $N$ sites follows the power-law $\bar C \sim N^{-c_0}$ with $c_0 = 0.7$ for $k_\mathrm{min} = 1$. Surprisingly the exponent $c_0 = 0.24$ is found consistently for networks with different $k_\mathrm{min} > 1$, as long as $k_\mathrm{min} \ll N - 1$. Although the clustering coefficient decreases with system size, it decreases slower than for random networks or Barabasi-Albert networks, which are characterized by an exponent $c_0 = 1$ and  $c_0 \approx 0.75$ respectively.\\
For the same networks, the average shortest path length $\langle l \rangle$ follows the double logarithmic behavior $l = c_0 \ln (\ln (c_1 N)) + c_2$ for all $k_\mathrm{min} > 1$ with different fitting parameters $c_0, c_1$ and $c_2$. Notice that for $k_\mathrm{min} = 1$ the average shortest path length is not well defined, since the network consists of isolated clusters. The increase is slower than logarithmic, which is found in the random case. This dependence is typical for ultra-small networks \cite{cohen03}.\\
\section{Conclusions}
In summary we have introduced a new model to create scale-free networks based on
preferential depletion of edges (for $\alpha = 2$). The networks exhibit an exponent
$\gamma \approx 5/3$ for the degree distribution independent of parameters and
smaller than the values obtained by growth models. Interestingly this value is in
very close agreement with the corresponding exponent of the Escherichia coli
metabolic network \cite{Friedman} or the gene functional interaction network
\cite{Gu}. As opposed to these models, the depletion model can also generate
disconnected structures, as observed in many biological systems, by tuning the value
of the minimal number of edges. These results suggest that depletion rather than
growth is the mechanism at the basis of the emergence of scale-free networks in
biology.\\
\section*{Acknowledgments}
We acknowledge financial support from the ETH Competence Center 'Coping with Crises
in Complex Socio-Economic Systems' (CCSS) through ETH Research Grant CH1-01-08-2 and
FUNCAP.

\end{document}